\documentclass[]{aa}

\usepackage{graphics}
\usepackage{times}

\begin{document}

\title{The broadening of Fe~II lines by neutral hydrogen collisions
\thanks{Tables 1 and 2 are only available in electronic form at the CDS via anonymous ftp to cdsarc.u-strsbg.fr (130.79.125.5) or via http://cdsweb.u-strasbg.fr/Abstract.html}
}

\author{P.S. Barklem \and J. Aspelund-Johansson}
 
\offprints{P. S. Barklem,
\email{barklem@astro.uu.se}}

\institute{Department of Astronomy and Space Physics, Uppsala University, Box 515, S 751-20 Uppsala, Sweden}

\date{Received 2 December 2004 / Accepted 20 January 2005}

\abstract{  
Data for the broadening of 24188 Fe~II lines by collisions with neutral hydrogen atoms have been computed using the theory of Anstee \& O'Mara as extended to singly ionised species and higher orbital angular momentum states by Barklem \& O'Mara.  Data have been computed for all Fe~II lines between observed energy levels in the line lists of Kurucz with $\log gf > -5$ for which the theory is applicable.  The variable energy debt parameter $E_p$ used in computing the second order perturbation theory potential is chosen to be consistent with the long range dispersion interaction constant $C_6$ computed using the $f$-values from Kurucz.  
}

\maketitle

\section{Introduction}
\label{sect:intro}

The line broadening theory of Anstee \& O'Mara (\cite{anstee91, anstee95}) was developed for $s$--$p$ and $p$--$s$ transitions between low-lying states of a neutral atom by neutral hydrogen collisions, with the hope of being able to compute data for a large number of lines of various elements for astrophysical applications.  The approximations necessary for development of such a generally applicable theory led to an expectation that the theory might enable cross-sections to be calculated with accuracies of around 20\%; a significant improvement on the classical van der Waals description (Uns\"old~\cite{unsold55}), commonly used in stellar spectroscopy due to its general applicability, with uncertainties of typically a factor of 2 or 3.  Application of the data to solar lines of a number of elements, including Na, Ca and Fe (Anstee \& O'Mara \cite{anstee95}, Anstee et al.~\cite{anstee97}), in fact suggested that the line broadening cross-sections might have uncertainties as low as 5\% for the considered lines, as judged from the internal consistency and agreement of abundances with the meteoritic abundances.  Anstee \& O'Mara (\cite{anstee91, anstee95}) suggested that this may be the result of error cancellation.  Recent comparisons with detailed calculations (eg. Kerkeni et al.~\cite{kerkeni04}, Barklem \& O'Mara \cite{barklem01}) indicate uncertainties in the range 5--20\%, the largest differences occuring in lines where the influence of avoided ionic crossings turns out to be particularly important in the adiabatic interaction potential, an effect which is neglected in the Anstee \& O'Mara theory to retain generalility.  The theory has been extended to transitions between higher orbital angular momentum states (Barklem \& O'Mara \cite{barklem97}, Barklem et~al. \cite{barklem98a}) and to the case of a singly ionized atom (Barklem \& O'Mara \cite{barklem98b}).  We will refer to these collected works by Anstee, Barklem and O'Mara as ABO theory. The extended theories have also been tested against the solar spectrum where possible and there is no evidence that these calculations have uncertainties significantly different from the earlier calculations.  

The theory for ionized atoms differs significantly from the theory for neutral atoms, in that calculations must be done line by line.  In the ABO theory, where potentials are computed using Rayleigh-Schr\"odinger perturbation theory, a key parameter is $E_p$, the fixed energy debt in the Uns\"old~(\cite{unsold27}) approximation to the second order interaction energy between the perturbed absorbing atom and the ground state H atom.  For neutral atoms energy level splittings are typically small compared to those for the hydrogen ground state and thus hydrogenic contributions are expected dominate the long range interaction.  In this approximation, also due to Uns\"old~(\cite{unsold55}), $E_p\approx-4/9$ atomic units irrespective of the species or state of the neutral atom.  This approximation enabled general tables of line broadening data to be computed, which could be interpolated in for any species.  In ionized atoms, energy spacings are typically larger than those in neutral atoms, and thus the Uns\"old approximation of $E_p=-4/9$ is questionable.  Barklem \& O'Mara~(\cite{barklem98b}) showed that while for the considered lines of Ca II and Ba II this approximation was reasonable, for the resonance lines of Mg II it led to an error of over 30\%.  Such a large errror, when combined with the intrinsic error in the theory, is considered unacceptably large for astrophysical application.  Thus, for ionized atoms calculations would need to be done on a line to line basis.

Up to now, calculations for ionized atoms have only been performed for a small number of astrophysically important lines (Barklem \& O'Mara \cite{barklem98b, barklem00}) due to the need to choose a suitable value for $E_p$.  The value of $E_p$ may be inferred from the dispersion (i.e. van der Waals) component of the long range interaction $C_6/R^6$, by choosing $E_p$ such that the potential will have this long range behaviour, and assuming this value at all separations.  Calculation of $C_6$ involves a summation over all $f$-values connected to the states of the two atoms.  In these past works the $f$-values for the perturbed atom have being collected from the literature, making the procedure labourious.  For some species, an alternative approach to this problem is provided by the line lists computed by Kurucz using a semi-empirical least-squares approach to Slater-Condon theory (Kurucz~\cite{kurucz73}, Cowan~\cite{cowan} \S16-3).  While the $f$-values computed via this method are perhaps not as accurate as experimental values which might be found in the literature for a small number of selected lines, the enormous advantage is an unparalleled degree of completeness which is important for our problem.  As we will discuss, the accuracy should be sufficient for our purposes.  The line lists provide not only the required $f$-values for computing $C_6$, but, of course, an extensive list of lines for which we can compute line broadening cross-sections.  We have computed such data for Fe~II, and the calculations and results are now described.

\section{Line Broadening Calculations}
\label{sect:theory}

The theory, particularly the potentials and collision dynamics, has been described in the earlier ABO theory papers.  For our specific case of a singly ionised Fe atom, the theory is described particularly by Barklem \& O'Mara (\cite{barklem98b}, hereafter BO98, and references therein).   

To perform the calculations we have used information from three files from Kurucz (\cite{kurucz03}; kurucz.harvard.edu): the Fe~II energy level list (gf2601.gam), the complete line list (gf2601.lin), and the line list (gf2601.pos) which includes only lines between observed levels.  These files were computed in 2003, following the method earlier described by Kurucz~(\cite{kurucz73}, \cite{kurucz81}) but with updated observed energy levels and identifications.

First, the various input parameters for the ABO theory must be adopted for each energy level in the Kurucz energy level list (gf2601.gam).  The ABO model for the interaction potential describes the absorbing atom as a positively charged core with a single optical electron.  The unperturbed optical electron is considered as moving independently of the core electrons in a central field, and thus its state may be characterised by principal quantum number $n$ and orbital angular momentum quantum $l$ along with its binding energy.  Together, these three pieces of information allow the unperturbed wavefunction for the optical electron to be computed in this approximation.  Such a description is justified by the fact that the broadening is dominated by separations on the tail of the electronic wavefunction (Anstee \&\ O'Mara~\cite{anstee91}).
   
For each level we determine this information from the dominant electron configuration and term provided in the Kurucz energy level list.  In all cases we assume the electron (or one of the electrons) with the highest $l$ of those electrons with the highest $n$ in the dominant configuration is the optical electron, since this electron will have the most extended wavefunction.  The binding energy of the electron is given by the difference between the energy of the state $E$ and the series limit $E_\mathrm{limit}$ which defines the Fe~III parent state.  The idea of parentage is only clear in the case of, like our model, a single running electron outside a more tightly bound core.  In more complex situations, particularly where there are multiple open shells, one often has parental mixing (see eg.\ Condon \&\ Odabasi~\cite{condon_odabasi}, Cowan~\cite{cowan}).  

The most common electron configuration in Fe II is of the form $3d^6(\alpha^\prime {}^{M_p}L_p) nl \;\alpha {}^ML$ where $\alpha^\prime {}^{M_p}L_p$ defines clearly the parent term in Fe~III.  However, in other configurations the parentage may be unclear or parental mixing occurs.  Levels of the form $3d^5 (\alpha^\prime {}^{M_g}L_g) 4s4p \;\alpha {}^ML$, where $\alpha^\prime {}^{M_g}L_g$ defines the grandparent, assuming the $4p$ electron to be our optical electron, has the possible parents ${}^{M_g\pm1}L_g$.  If $M=M_g+2$ then we must have ${}^{M_g+1}L_g$, while if $M=M_g-2$ we must have ${}^{M_g-1}L_g$.  If $M=M_g$ then either possibility may occur and we arbitrarily chose the alternative giving the lowest binding energy, and thus the most extended wavefunction.  

While our model is not really applicable to the case where the external shell contains equivalent electrons, we proceed on the assumption that a reasonable estimate might be obtained by taking one of the equivalent electrons as the optical electron.  We note that since the $C_6$ calculation is not dependent on the optical electron model, our long range potential will be correct since it is referenced to the $C_6$ calculation.  For the configuration $3d^7$, which has a large number of fractional parents, we used the principal parent (see Condon \&\ Odabasi~\cite{condon_odabasi}), namely $a {}^3F$.  The configuration $3d^5 4s^2 \;\alpha {}^ML$ has two possible parents, namely ${}^{M\pm1}L$.  In an attempt to best model the wavefunction of the $4s$ electron with the most extended wavefunction we assigned the parent as that parent which gives the smallest binding energy which is ${}^{M+1}L$ by Hund's rule.    Finally, for the small number of levels of the form $3d^5 (\alpha^\prime {}^{M_g}L_g) 4p^2({}^1S,{}^3P,{}^1D) \;\alpha {}^ML$ treated in this work we have assumed that the parent is $z {}^7P$ unless $M=4$ in which case we assume $z {}^5P$. 

Having assigned to each state $n$, $l$ and parent terms, and thus binding energies for the optical electron, scaled Thomas-Fermi-Dirac radial wavefunctions (Warner~\cite{warner68}) were then computed for the optical electron for each state.  Due to neglect of exchange effects ABO calculations are only valid for low-lying states where such effects are not large (see relevant ABO papers), and thus we treat only states for which the effective principal quantum number $n^\star$ is less than 3 for $s$ and $p$ states and for which $n^\star < 4$ for $d$ states.

Now we need to specify $E_p$ for each state.  For a perturbed atom $A$ with states $k^\prime$ in given energy level $k$, interacting with a H atom in the $1s$ ground state, denoted by $l$, with excited states $l^\prime$, the $C_6$ value is computed via the standard expression, in atomic units,  
\begin{equation}
C_6 = \frac{3}{2} \sum_{k^\prime\neq k}\sum_{l^\prime\neq l}
\frac{f^A_{kk^\prime} f^H_{ll^\prime}}
{(\Delta E_{k^\prime k}^A + \Delta E_{l^\prime l}^H)
\Delta E_{k^\prime k}^A  \Delta E_{l^\prime l}^H},
\label{eqn:C6def}
\end{equation}
where $\Delta E^X_{m^\prime m} = E_{m^\prime}^X - E^X_m$.  Data for bound-bound transitions in the Fe~II atom are taken from the Kurucz complete line list (gf2601.lin).  The bound-bound transition data for H are taken from Wiese, Smith and Glennon~(\cite{wiese66}).   
As in BO98, bound-free transitions are accounted for in the H atom, but not in the perturbed atom.  In BO98 an approximate constant Gaunt factor for the bound-free component was adopted; a polynomial fit to the calculations of Karzas \&\ Latter (\cite{karzas61}) is now used.  Once the $C_6$ coefficient is computed, the value of $E_p$ may be chosen such that the ABO theory Rayleigh-Schr\"odinger-Uns\"old potential will have the correct long range behaviour via \begin{equation}
E_p = - \frac{2 \langle p_2^2 \rangle }{C_6}.
\label{eqn:Eprelation}
\end{equation} 
The mean square position of the optical electron $\langle p_2^2 \rangle$ is computed from the wavefunction.  

Kurucz has also computed $C_6$ values, provided in his energy level list (gf2601.gam), making the approximation that the energy levels in the perturbed atom are much more closely spaced than the separation of the H ground state from other H states (Kurucz~\cite{kurucz81}, page 76), which leads to
\begin{equation}
C_6 = \frac{3}{2} \alpha_{\mathrm{H}(1s)} \sum_{k^\prime\neq k}
\frac{f^A_{kk^\prime}}
{\Delta E_{k^\prime k}^A },
\label{eqn:C6def_app}
\end{equation}
where $\alpha_{\mathrm{H}(1s)}$ is the static dipole polarizability of the H atom in the ground state, which is $9/2$ atomic units (Dalgarno~\cite{dalgarno62}).  As a check on our calculations, we also computed $C_6$ in this approximation as well as the $f$-value sum. The results were in good agreement with those computed by Kurucz.  Comparison of our $C_6$ values from equation~\ref{eqn:C6def} with those computed by Kurucz indicates that this approximation leads to overestimation of $C_6$ by a mean factor of 1.23, with a standard deviation of 0.47, in Fe~II. 

The results for $C_6$ and $E_p$ are presented in Table~\ref{tab:levels} as they may be of interest for other applications, for example calculation of collisional depolarization cross-sections.  The static dipole polarizabilities are presented since they are easily computed and such fundamental data may be of interest, for example, in estimating long range interactions with perturbers other than H.  Note that we consider only observed energy levels (i.e.\ no predicted levels) and states which meet the criteria for $n^*$ discussed above for the applicability of the ABO theory.

\begin{table*}
\tabcolsep 1.5mm
\begin{center}
\caption{The computed interaction parameters for the 572 considered levels of Fe II.   The complete table is available only electronically; a short extract is provided here as a guide to its form.  For each level tabulated are: the energy $E$ with respect to the ground state, the parity (EVE or ODD), the $J$ quantum number, the principal quantum number $n$ for the optical electron, the orbital angular momentum quantum number $l$ for the optical electron, the adopted series limit energy $E^\mathrm{limit}$, the mean square position of the optical electron $\langle p_2^2 \rangle$, the dispersion coefficient $C_6$ and the static dipole polarizability $\alpha_d$ of the level, and the energy denominator $E_p$.  Note that we use the definition $\Delta E = C_6/R^6$, while Kurucz uses $\Delta \nu = C_6^\mathrm{KURUCZ}/R^6$; thus the definitions differ by a factor of $h$, in addition to the difference in units.}
\label{tab:levels}
\small
\begin{tabular}{rccccccccc}
\hline
$E$ & Parity & $J$ & $n$ & $l$ & $E^\mathrm{limit}$ & $\langle p_2^2 \rangle$ & $C_6$ & $\alpha_d$ & $E_p$ \\  
{[cm$^{-1}$]} & & & & & [cm$^{-1}$] & [a.u.] & [a.u.] & [a.u.] & [a.u.] \\  
\hline 
       0.000 &   EVE & 4.5 & 4 & 0 & 130563.000   &   8.53  &  29.73  &   30.5 &  $-0.574$ \\
     384.790 &   EVE & 3.5 & 4 & 0 & 130563.000   &   8.57  &  29.73  &   30.6 &  $-0.577$ \\
$\vdots$  & $\vdots$ & $\vdots$ & $\vdots$ & $\vdots$ & $\vdots$ & $\vdots$ & $\vdots$ & $\vdots$ & $\vdots$ \\ 
\hline
\end{tabular}
\end{center}
\end{table*}

Line broadening cross sections were then computed for lines in Kurucz's list of Fe~II lines (gf2601.pos) with $\log gf > -5$, where the upper and lower states are both in Table~\ref{tab:levels}.  Potentials are computed employing the computed wavefunctions and adopting the computed $E_p$ values.  We have limited the calculations as this mechanism is not important in very weak lines and computing times are non-negligible, of the order of a minute per line on a modern workstation.  Note that our calculations deal only with transitions where $\Delta l=\pm1$; lines not obeying this selection rule are typically weak in any case.          
\begin{table*}
\tabcolsep 1.5mm
\begin{center}
\caption{The computed broadening data.  The entire table is available only electronically; a short extract is provided here as a guide to its form.  The table presents for each line:  the species as described by the packed parameter $Z+0.01(Z-N)$ where $Z$ is the atomic number and $N$ is the number of electrons, the air wavelength $\lambda_\mathrm{air}$, the $J$ quantum numbers for the lower and upper states $J_\mathrm{low}$ and $J_\mathrm{upp}$, the energies of the lower and upper states $E_\mathrm{low}$ and $E_\mathrm{upp}$, the line broadening cross-section $\sigma$ for a collision speed of $10^4$~m~s$^{-1}$, the dimensionless velocity parameter $\alpha$, and the log of the line width (FWHM) per perturber at a temperature of $10^4$~K.  Further, the cross section and velocity parameter computed using the approximation $E_p=-4/9$ a.u. are given, $\sigma_{4/9}$ and $\alpha_{4/9}$.}
\label{tab:lines}
\small
\begin{tabular}{ccccccccccc}
\hline
Species & $\lambda_\mathrm{air}$ & $J_\mathrm{low}$ & $J_\mathrm{upp}$ &  $E_\mathrm{low}$ & $E_\mathrm{upp}$  &  $\sigma$ & $\alpha$ & $\log(\Gamma_\mathrm{10^4K}/N_H)$ & $\sigma_{4/9}$ & $\alpha_{4/9}$\\   
 & [\AA] & & &  [cm$^{-1}$] & [cm$^{-1}$] & [a.u.] & & [rad s$^{-1}$ cm$^3$]  & [a.u.] &\\   
\hline    
 26.01  & 890.122 &  4.5  &  5.5 &      0.000 & 112344.110  &  326. &  0.319 &   $-7.634$ &   163. &  0.200 \\
 26.01  & 905.210 &  4.5  &  5.5 &   1872.567 & 112344.110  &  312. &  0.437 &   $-7.674$ &   203. &  0.263 \\
 $\vdots$ & $\vdots$ & $\vdots$ & $\vdots$  & $\vdots$ & $\vdots$ & $\vdots$ & $\vdots$ & $\vdots$ & $\vdots$ & $\vdots$\\ 
\hline
\end{tabular}
\end{center}
\end{table*}

The computed line broadening data are presented in Table~\ref{tab:lines}.  We have included sufficient information for lines and states involved to be uniquely identified.  In table~\ref{tab:lines}, we quote the line broadening cross section $\sigma$ for a collision speed $v=10^4$~m~s$^{-1}$ and the velocity exponent $\alpha$ as defined by equation 1 of  Anstee \& O'Mara~(\cite{anstee95}).  They also provide an expression for calculation of the line half half-width per perturber for a Maxwellian velocity distribution from these data, their equation 3.  In table~\ref{tab:lines} we quote the line full half-width per unit hydrogen atom density for 10$^4$~K.  The temperature dependence of the line width is $T^\frac{1-\alpha}{2}$.  We also provide the line broadening data computed in the Uns\"old approximation $E_p=-4/9$, namely the cross section $\sigma_{4/9}$ and velocity parameter $\alpha_{4/9}$.  These data are of interest in estimating the sensitivity of the data to the value of $E_p$.

\section{Discussion}
\label{sect:discussion}

We now discuss the results including some statistical properties, their impact and reliability.  

\begin{figure}
\begin{center}
\resizebox{69mm}{!}{\rotatebox{0}{\includegraphics{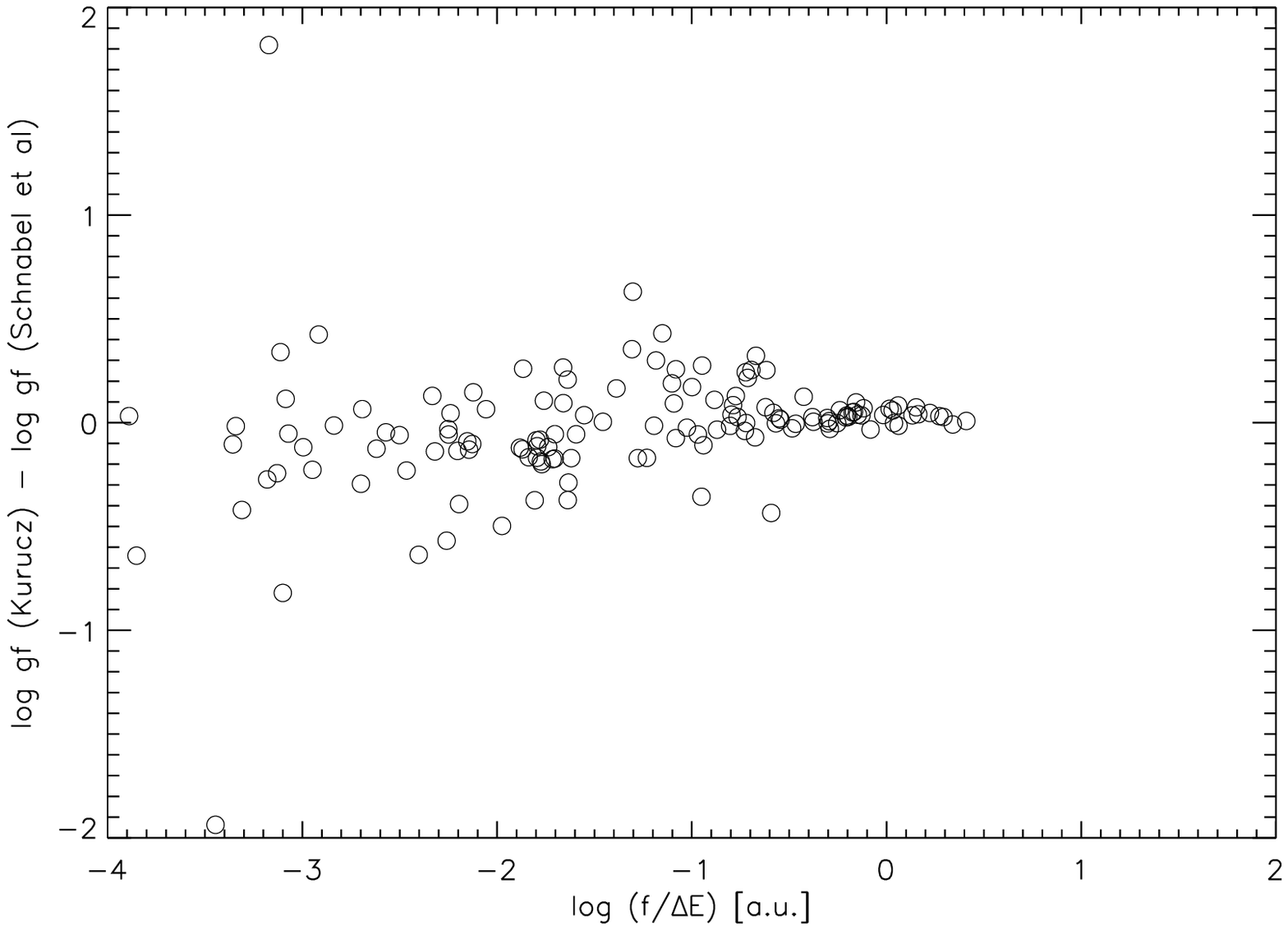}}}
\resizebox{69mm}{!}{\rotatebox{0}{\includegraphics{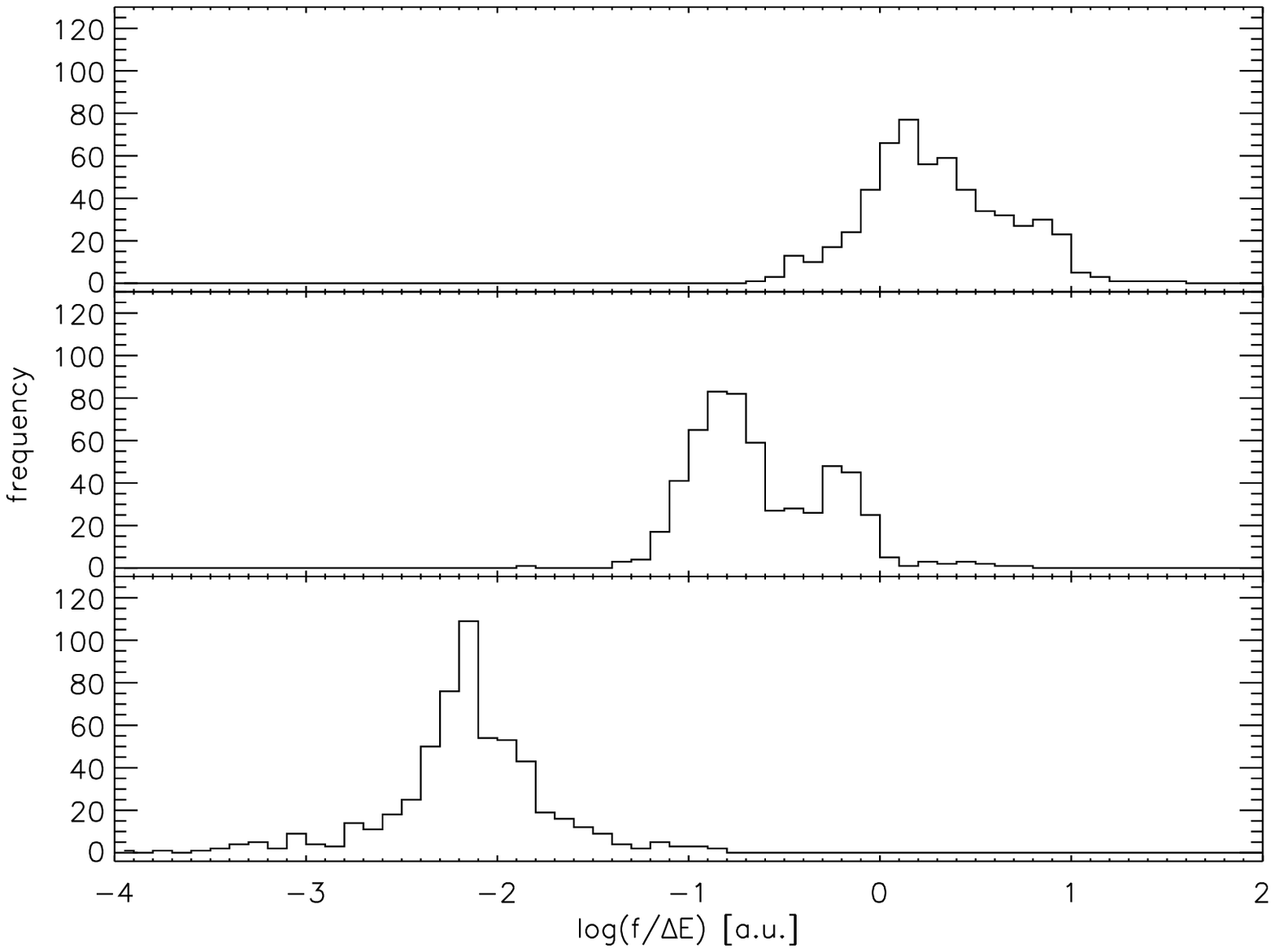}}}
\end{center}
\caption{The upper plot shows a comparison of $f$-values from Kurucz~(\cite{kurucz03}) with experimental data from Schnabel et~al.~(\cite{schnabel04}), plotted against the logarithm of the transition's approximate weight in a $C_6$ sum, $\log (f/\Delta E)$.  The lower plot shows histograms of $\log (f/\Delta E)$ for the most important transitions (upper panel), 10th most important (middle panel) and 100th most important (lower panel) transitions to each $C_6$ sum for all considered states in our calculations. }
\label{fig:f_comp}
\end{figure}

First, we discuss in general terms the accuracy and reliability of the calculations.  Assuming the completeness of the most important data for the $C_6$ sum, the accuracy of the computed $C_6$ will depend on the accuracy of the $f$-values for Fe~II, the data for H being essentially exact.  To estimate the accuracy of the Kurucz $f$-values we have compared them with the experimental data of Schnabel et~al.~(\cite{schnabel04}) in Fig.~\ref{fig:f_comp}. The experimental data have uncertainties ranging from 0.025 dex for the strongest lines, to 0.1 dex for the weakest lines. The comparison has been plotted against $f/\Delta E$, which reflects the weight the transition has in the $C_6$ sum (see equation~\ref{eqn:C6def} and~\ref{eqn:C6def_app}).  Overall the data compare well, with no evidence for any systematic differences within the uncertainty in the experimental data. The mean difference is $-$0.023 dex with a standard deviation of 0.31 dex.  However, it is clear that the stronger transitions are in the best agreement.  As shown in the lower panel of Fig.~\ref{fig:f_comp}, these are the most important transitions in contributing to the $C_6$ sum, and thus the error in these $f$-values is more relevant for estimating the error in $C_6$.  If we consider only those transitions with $\log (f/\Delta E) > -1$ the mean difference is $+$0.036 dex with a standard deviation of 0.12 dex.  The comparison suggests that the random error in the $f$-values for the most important transitions is about 0.1~dex. Thus, we estimate the error in the computed $C_6$ values at around this level, i.e.\ 26\%.   A rough estimate of the error in the broadening cross section can be made via the van der Waals theory, in which the cross section is proportional to $C_6^{2/5}$.  Thus, a 26\% uncertainty in $C_6$ leads roughly to an uncertainty of 10\% in the cross section.   

As seen in Fig.~\ref{fig:f_comp}, there are a small number of quite discrepant transitions.  For reasons pointed out by Kurucz~(\cite{kurucz81}, pg 29), as in any such large-scale calculation, data that are grossly in error may arise.  The expectation is, as seen in the Fig.~\ref{fig:f_comp}, that these discrepancies are more likely for weak, and thus less important, transitions.  We cannot, however, rule out such large errors affecting important transitions in our calculations.  It should also be mentioned that our neglect of bound-free transitions in the Fe~II atom could be important for some higher states. 

\begin{figure}
\begin{center}
\resizebox{69mm}{!}{\rotatebox{0}{\includegraphics{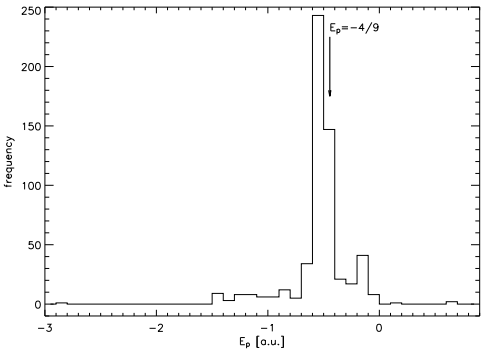}}}
\end{center}
\caption{Histogram of $E_p$ values.  The Uns\"old approximation value of $E_p=-4/9$ atomic units is indicated.}
\label{fig:ep_levels}
\end{figure}

\begin{figure}
\begin{center}
\resizebox{69mm}{!}{\rotatebox{0}{\includegraphics{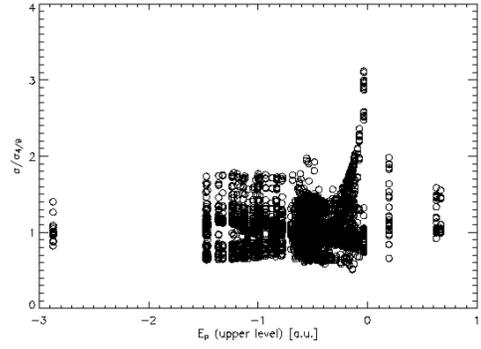}}}
\end{center}
\caption{Plot of the ratio $\sigma/\sigma_{4/9}$ with the $E_p$ for the upper state.}
\label{fig:ratio_cross49}
\end{figure}

\begin{figure}
\begin{center}
\resizebox{69mm}{!}{\rotatebox{0}{\includegraphics{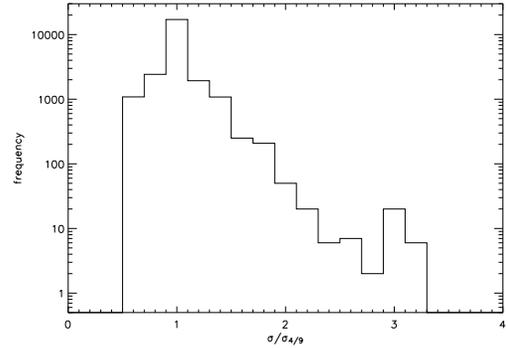}}}
\end{center}
\caption{Histogram of the ratio $\sigma/\sigma_{4/9}$.}
\label{fig:ratio_49}
\end{figure}

Figure~\ref{fig:ep_levels} plots the distribution of the inferred $E_p$ values for the considered states.  We see that the majority of states have $E_p$ values close to Uns\"old's hydrogenic value of $-4/9$ atomic units.  Notably, there are a number of states for which a low value of $|E_p|$ is inferred.  As the second order dispersive interaction is inversely proportional to $E_p$, the interaction energy, and thus the line broadening cross section, are particularly sensitive to the choice of $E_p$ value when its absolute value is small (see Fig.~1 of Barklem \& O'Mara~\cite{barklem00}).  Figure~\ref{fig:ratio_cross49} plots the ratio $\sigma/\sigma_{4/9}$ against the $E_p$ value for the upper state.  It demonstrates that the breakdown of the $E_p=-4/9$ approximation can be important when $|E_p|$ is small.  We note that the states with small $|E_p|$ often correspond to states where the external shell contains equivalent electrons.  Small $|E_p|$ values arise in such cases due to small $\langle p_2^2 \rangle$ values (see equation~\ref{eqn:Eprelation}) resulting from the compact wavefunction for the chosen optical electron in such states.  Note that Fig.~\ref{fig:ratio_cross49} is somewhat deceptive in depicting the scatter in this ratio, and thus a corresponding histogram of $\sigma/\sigma_{4/9}$ is given in Fig.~\ref{fig:ratio_49} which will be discussed below.

\begin{figure}
\begin{center}
\resizebox{69mm}{!}{\rotatebox{0}{\includegraphics{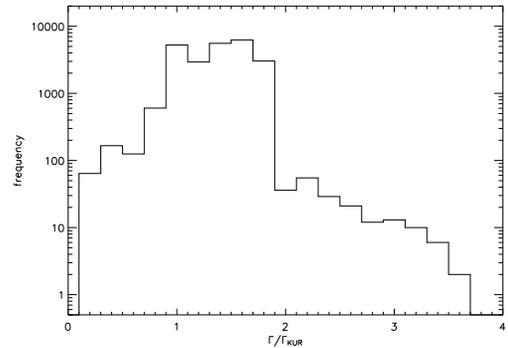}}}
\end{center}
\caption{Histogram of the line width ratios between this work and van der Waals theory as computed by Kurucz.}
\label{fig:ratio_vdw}
\end{figure}

Figure~\ref{fig:ratio_vdw} compares our line width results with those computed by Kurucz using the classical van der Waals broadening theory (Uns\"old~\cite{unsold55}).  Our line widths are typically larger, with a mean enhancement factor of 1.37 with a standard deviation of 0.31.  Figure~\ref{fig:ratio_49} compares our best cross sections $\sigma$ with those computed in the Uns\"old approximation $\sigma_{4/9}$.  As expected given that most levels have an $E_p$ near this value (see Fig.~\ref{fig:ep_levels}), for the vast majority of lines the results are in good agreement.  However, there are a number of lines where the approximation breaks down.  As shown in Fig.~\ref{fig:ratio_cross49}, these cases correspond to lines with small $|E_p|$ values.

\section{Concluding remarks}
\label{sect:conclusion}

We have computed line broadening data for 24188 lines of Fe~II from the Kurucz line lists with $\log gf > -5$ where the ABO theory is applicable.  Thus, we have demonstrated the possibility for such large-scale calculations using the theory.  Similar calculations should be done for other species, including neutral species.  Such data will be of importance in both interpreting individual lines in cool star spectra and in opacity calculations for modelling cool star atmospheres.   The influence of collisional broadening due to hydrogen on the total line blanketing and line blocking can be important (eg. Short \& Hauschildt~\cite{short04}).
  
We estimate the uncertainty in the computations of the dispersion coefficient $C_6$ as approximately 26\% leading to an uncertainty of 10\% in the cross section (\S~3).  Based on our discussion in section~1, the intrinsic uncertainty for the theory is around 10\%.  Thus, we estimate that the uncertainty in the calculations is around 20\%.  Occasional large errors may occur.  We plan to do similar calculations for all neutral and singly ionized species for which Kurucz has computed data.  The data will be incorporated into the Vienna Atomic Line Database (VALD, Kupka et~al.~\cite{vald}).   

\begin{acknowledgements}

We gratefully acknowledge the support of the Swedish Research Council. 

\end{acknowledgements}

\end{document}